# MID-TERM REVIEW OF THE IAU STRATEGIC PLAN 2020-2030
Debra Elmegreen, IAU President
June 2024

## I. INTRODUCTION: IAU STRATEGIC PLAN 2020-2030

At the XXX[th] General Assembly in Vienna in 2018, Resolution A1 called for a mid-term review of the IAU Strategic Plan 2020-2030 in time for the General Assembly in 2024. In particular, the resolution states in part:

### Resolution A1 at the XXX[th] General Assembly in Vienna, 2018
*"11. That, starting in 2019, the Executive Committee should prepare the implementation of the Strategic Plan 2020-2030...*
*12. That the goals of the Strategic Plan 2020-2030 should be pursued to their full extent starting in 2021.*
*13. That the Executive Committee shall include mid-term reviews about the implementation of the Strategic Plan 2020-2030 in the EC Triennial Reports to be presented to the XXXII[nd] and XXXIII[rd] General Assemblies in 2024 and 2027.*
*14. That a new Strategic Plan 2030-2040 should be prepared by the Executive Committee to be presented for approval to the XXXIV[th] General Assembly in 2030."*

This report constitutes the 2024 mid-term review of the Strategic Plan, which was presented at the 110[th] Executive Committee meeting in April 2024. To place the review in context, first recall the five primary goals from the Strategic Plan, listed below:

### GOALS FROM THE IAU STRATEGIC PLAN 2020-2030

**Goal 1**
*The IAU leads the worldwide coordination of astronomy and the fostering of communication and dissemination of astronomical knowledge among professional astronomers.*
**Goal 2**
*The IAU promotes the inclusive advancement of the field of astronomy in every country.*
**Goal 3**
*The IAU promotes the use of astronomy as a tool for development in every country.*
**Goal 4**
*The IAU engages the public in astronomy through access to astronomical information and communication of the science of astronomy.*
**Goal 5**
*The IAU stimulates the use of astronomy for teaching and education at school level.*

## II. ACTIONS IN THE STRATEGIC PLAN

The Strategic Plan includes a description of many longstanding activities, in addition to specific new actions that are intended to help achieve each primary goal. The ongoing activities are not presented in detail here, but links are given for their reports and reviews. This review focuses on the new actions in the Strategic Plan, listed below in italics in the order presented in the Plan, with a summary of what has been accomplished so far for each of them.

### GOAL 1 ACTIONS

Goal 1 primarily pertains to the general scientific efforts of the IAU. The IAU's efforts on the advancement of astronomy through coordination and communication take place through the 9 scientific Divisions, 39 Commissions, and 48 Working Groups, in addition to IAU meetings. There are typically 9 IAU Symposia each year, and 3 Regional Meetings during a triennium. The General Assembly is held every 3 years, which includes Symposia, Focus Meetings, Division Days, Sessions for the Offices and the Executive Committee Working Groups (ECWGs), and plenary, prize, and public talks.

As part of these efforts:
- *"The IAU will expand membership to include the category of Junior Members for early career astronomers, and will stimulate activities fostering their professional development."*
  **Efforts to date:**
  The Junior Member category that was started in 2018 surpassed 1200 members in 2023. An ECWG on Junior Members was formed. It has hosted many online webinars as well as sessions at the General Assembly, focused on professional development, mentoring, networking, mental health, and careers.
  In addition, the IAU Hands-on Workshops (I-HOW), funded by the Gordon and Betty Moore Foundation since 2022, help provide training for early career astronomers, primarily in less developed regions. I-HOWs have been held so far in Santiago, Chile on VLT High Angular Resolution Observations in October 2022, jointly with COSPAR in Potchefsroom, South Africa on X-ray Vision of our Universe in February 2023, and a joint Iran-Turkey workshop on Radio Astronomy in Turkey, September 2023. I-HOWs are planned in Thailand on JWST astronomy and in China on X-ray spectroscopy in summer 2024, and plans are underway for I-HOWs in UAE on optical astronomy and Poland on JWST astronomy in 2025.
  Training is also available through the Office for Young Astronomers (OYA) and their International Schools for Young Astronomers (ISYA), discussed further in Goal 2 actions below. As part of its efforts, the OYA also coordinates with other IAU scientific bodies and Offices and provides guidance on diverse career trajectories both within and beyond academia, provides mentorship in isolated areas of the world, and promotes the use of new methods of learning and new technologies.

- *"The EC will continue IAU-related publications and communications, using and promoting updated methods."*
  **Efforts to date:**
  The IAU publishes *Proceedings* of IAU Symposia, *Astronomy in Focus* on General Assembly Focus Meetings, *Transactions* on the business and ceremonies of the General Assembly, biannual *Catalyst* newsletters on IAU activities, and the journal on *Communicating Astronomy to the Public (CAP)*, in addition to brochures and material available through its Offices and Center websites. Division, Commission, and Working Group reports are also listed.
  The Secretariat also produces and posts annual reports and brochures, and makes use of social media (X, LinkedIn, Facebook). In addition, the General Secretary sends out newsletters to IAU members after press releases, and the President sends bimonthly email messages to IAU members.

- *"The Working Group on Global Coordination will have a focus meeting at the GAs and a focused workshop between GAs."*
  **Efforts to date:**
  The ECWG on Global Coordination has held workshops including future space-based UVOIR telescopes, Leiden 2017; multimessenger transient astronomy, Cape Town 2020; and future far-IR to mm astronomy, Pasadena 2024. These were funded by the Kavli Foundation, and will continue as funding allows. The ECWG also hosts sessions at the GAs to facilitate discussions on coordination of efforts.

- *"The IAU will organise an international conference on dark and quiet sky protection, in cooperation with COPUOS and UNESCO, as well as follow-up meetings and interactions with IUCN, to raise awareness at all levels."*
  **Efforts to date:**
  Conferences hosted jointly by the IAU, UNOOSA, and Spain were held online for Dark & Quiet Skies I, Oct. 2020, and II, Oct. 2021. IAU Symposium IAUS385: Astronomy and Satellite Constellations: Pathways Forward was in Canary Islands in 2023, and IAUS386: Dark Sky and Astronomical Heritage in Boosting Astro-tourism Around the Globe was in Ethiopia in 2023. The ECWG on Dark and Quiet Sky Protection and its associated Commissions and Working Groups continue with technical and policy efforts.
  The Center for the Protection of the Dark & Quiet Sky from Satellite Constellation Interference (CPS) was formed in April 2022, cohosted by NSF's NOIRLab and the SKA Observatory. It constitutes 4 main hubs (Sathub, Industry and Technology, Community Engagement, and Policy) with hundreds of contributing and affiliated members. The CPS represents the IAU and dark and quiet sky efforts at COPUOS meetings, supported by a Group of Friends from several countries. The CPS produced a recommendations paper in 2024, and submitted a resolution on the protection of the dark and quiet sky from harmful interference by satellite constellations for GA2024.

- *"The IAU will continue to foster connections with relevant professional organisations to help advance science."*
  **Efforts to date:**
  IAU has representatives on over 2 dozen [external committees](). It is especially involved in the International Science Council ([ISC]()) and its GeoUnions, with the current IAU President on the ISC governance advisory committee, and the President-elect as the IAU representative to the ISC. The current IAU President also serves on the U.S. Board on International Science Organizations, which encompasses 18 scientific unions.
  The IAU has MoUs with the Committee on Space Research (COSPAR), the International Association of Physics Students ({iaps}), the International Year of Basic Sciences for Sustainable Development (IYBSSD), the Activation Phase of The Earth-Humanity Coalition: Sciences for Equitable Wellbeing on a Healthy Planet Enabling the International Decade of Sciences for Sustainable Development 2024-2033 (IDSSD), and the International Congress of the Astronomical Youth, and has helped sponsor several non-IAU regional meetings in Africa and elsewhere.

- *"The IAU will continue to oversee official assigning of names for celestial bodies and their features."*
  **Efforts to date:**
  There is continued naming oversight via [ECWGs]() on Exoplanetary System Nomenclature ([WGESN]()), Planetary System Nomenclature ([WGPSN]()) and Small Bodies Nomenclature ([WGSBN]()). During the 2021-2024 triennium, the WGESN, formed in 2021, proposed guidelines for naming conventions for exoplanets and their host stars. With OAO, it oversaw the [NameExoWorlds]() 2022 Campaign (following the 2019 [NameExoWorlds]() competition during [IAU100]()) to name 20 pairs of exoplanets and their host stars targeted by JWST. The WGPSN approved 237 new names for planetary surface features and 13 names for moons of Saturn, in addition to other actions. The WGSBN publishes monthly [Bulletins](), each featuring dozens of new names for asteroids and comets, and collaborated on a [public contest]() to name a quasi-moon of Earth.

- *"The IAU will continue to oversee the definition, determination, and use of astronomical standards."*
  **Efforts to date:**
  Efforts on astronomical standards continue via specific [Division A Working Groups]().
  Two [resolutions]() on lunar time standards were submitted for the 2024 GA: one on a standard Lunar Celestial Reference System and Lunar Coordinate Time, and another to establish a coordinated lunar time standard.

- *"The IAU will continue to enhance and diversify its portfolio of prizes to reflect changes in its priorities."*
  **Efforts to date:**
  Prizes for Outreach, Development, and Education ([ODE]()) efforts were initiated in 2022 and are awarded triennially at GAs.
  Divisions have awarded annual [PhD]() Prizes and Honorable Mentions since 2018.

## GOAL 2 ACTIONS

Goal 2 pertains to many IAU-related efforts, both among the scientific bodies and among the Offices and Centers. The IAU strives for a gender and geographical balance of Executive Committee leaders and encourages balance for Division leaders and speakers and Scientific Organizing Committees for IAU Symposium and Focus Meetings.

In addition:
- *"The IAU will continue to increase the level of actions to promote diversity, encourage and retain women and minorities in astronomy, and support astronomers with special needs."*
  **Efforts to date:**
  Ongoing efforts continue via the ECWGs on Women in Astronomy (WIA) and Astronomy for Diversity & Inclusion (DI), the ISCU (now ISC) Gender Gap in Science report, the OAO WIA projects, Heising-Simons funding for Women & Girls in Astronomy, IAU100 activities and projects, and the first-ever IAU symposium on Equity, Diversity and Inclusion (IAUS 358), which resulted in a Springboard to Action report with recommendations.
  DI efforts also include the Inspiring Stars traveling program during IAU100 providing support and activities for the visually impaired, and the publication in 2021 of the second international comparative list of astronomical terms in sign languages.

- *"WIA and ICSU will cooperate on the Gender Gap in Science study and similar future studies to promote best practices for achieving gender parity."*
  **Efforts to date:**
  The ISCU (now ISC) Gender Gap study in which the WIA participated is complete; the IAU is open to participating in possible future studies.

- *"IAU-sanctioned events will follow anti-harassment guidelines."*
  **Efforts to date:**
  The IAU Code of Conduct includes an Ethics Policy and Anti-harassment Policy that apply to attendees at all IAU-related events. The Code was last updated in September 2023.

- *"The OYA (Office for Young Astronomers) will hold as many as 4 ISYA (International Schools for Young Astronomers) schools every 2-year period."*
  **Efforts to date:**
  ISYAs have been operating since 1967, and the OYA was founded in 2015 as a partnership between the IAU and the Norwegian Academy of Science and Letters to oversee the schools. The pandemic disrupted the schools during 2020-2022, but ISYAs resumed in Mexico and South Africa in 2023. In the immediate future, the OYA will continue with 3 ISYAs rather than 4 in every 2-year period, due to limited resources and manpower. The successes of the ISYAs and their participants are being archived to document and follow up on the half-century of engagement of the IAU with young astronomers.

**GOAL 3 ACTIONS**

Goal 3 primarily pertains to the Office of Astronomy for Development (OAD), founded in 2011 as a partnership between the IAU and the South African National Research Foundation and based at the South African Astronomical Observatory, with the support of the Department of Science and Innovation. The OAD has its own detailed plan, which complements the overall goals of the IAU. An external review of the OAD was completed in 2021, with a recommendation for it to continue.

In the context of the IAU Strategic Plan for 2020-2030, the OAD should:
- *"Contribute significantly to at least half of all United Nations Sustainable Development Goal (SDG) indicators; develop a number of global OAD "signature" projects."*
  **Efforts to date:**
  The OAD has developed 3 Flagship ("signature") themes: Socio-economic Development through Astronomy (Astrotourism), Astronomy for Mental Health, and Astronomy Knowledge and Skills for Development (Hack4Dev).
  Since its founding, the OAD has funded 220 projects in 108 countries and 5 continents that fall within the Flagship programs, which have impacted 2 million people and have contributed to at least 12 of the 17 SDGs. These include: #3 good health and well-being, #4 quality education, #5 gender equality, #6 clean water and sanitation, #7 affordable and clean energy, #8 decent work and economic growth, #9 industry, innovation and infrastructure, #10 reduced inequalities, #11 sustainable cities and communities, #13 climate action, #16 peace, justice and strong institutions, and #17 partnerships for the goals.

- *"Establish enough regional offices to cover all populated regions of the world."*
  **Efforts to date:**
  The OAD has established 11 Regional Offices and Language Centers so far, including in Europe (E-ROAD), founded in 2018, and in North America (NA-ROAD), founded in 2020.
  Large regions without an IAU Regional Office include Argentina (but it is reported that they are working with other groups, and they have hosted some OAD projects), India and Russia (which have also hosted OAD projects), and Australia (which is sparsely populated).

- "Refine the OAD project evaluation and feedback loop."
  **Efforts to date:**
  Ongoing, including advice to proposers from Regional Offices on how to write and improve a proposal, an independent review panel, and expert external consultation for project assessment when necessary.

- *"Use astronomy and its technology to position young people for career opportunities throughout society."*
  **Efforts to date:**
  Ongoing through multiple [projects](), including hackathons and coding, distance learning, tools for tracing health issues in a community, providing telescopes, mentoring, workshops, math skills, astronomy nights.

- *"Establish interdisciplinary partnerships around science for development."*
  **Efforts to date:**
  Interdisciplinary partnerships are an intrinsic part of various projects and Flagship themes. Mental health efforts include partnering with social workers, medical professionals, and both governmental and non-governmental collaborators. Hackathon projects include partnering with data science organizations and research groups. Astrotourism efforts involve collaborating with local businesses.

- *"Source the necessary funding to realise the above and assist other related initiatives in fundraising."*
  **Efforts to date:**
  Ongoing. A part-time fundraiser was hired to work with a volunteer fundraiser and AfAS for IAU-related activities in Africa. There is also a plan to develop video training for tips on fundraising and development.

## GOAL 4 ACTIONS

Goal 4 primarily pertains to the Office for Astronomy Outreach ([OAO]()), founded in 2012 as a joint partnership between the IAU and the National Astronomical Observatory of Japan (NAOJ) and based at NAOJ. The OAO has its own detailed plan, which complements the overall goals of IAU. An [external review]() of the OAO was completed in Fall 2023, with a recommendation for it to continue.

In the context of the IAU Strategic Plan for 2020-2030, the OAO should:
- *"Increase the network of National Outreach Coordinators (NOCs); restructure and ensure their effectiveness."*
  **Efforts to date:**
  There are now [NOCs]() in over 120 countries. There was a big push to recruit NOCs leading up to IAU100, with worldwide involvement in IAU100 events. More NOCs are being recruited in many regions, including recently in the U.S.

- *"Facilitate international communication through exchanges and translations."*
  **Efforts to date:**
  An Astronomy Translation Network was established to centralize resources that need translating and to facilitate volunteer participation in the effort, resulting in astronomical [glossaries]() in several languages.

Several international exoplanet naming contests, as well as a campaign to name a quasi-moon of Earth, have increased awareness of astronomy by engaging the public and schoolchildren.

- *"Provide open databases and public-friendly access to astronomical information."*
**Efforts to date:**
Basic information on various astronomical topics is available to the public. Many of the resources and publications are translated into multiple languages.

- *"Encourage communication of science and critical thinking through IAU member public engagement, professional-amateur, and citizen science activities."*
**Efforts to date:**
OAO publications, including monthly newsletters, the CAP journal, IAU Outreach Initiatives, and brochures on light pollution and on technical applications of astronomy to society, all encourage public awareness.

The Meet the Astronomers program facilitates meetings and talks between IAU members and amateur astronomy groups, the public, and students.

Some funded OAO projects involve citizen science activities.

The OAO staff are involved in the Executive Committee Working Group on Professional-Amateur Relations in Astronomy, which was established in April 2021. The WG held a session at the GA in Busan in 2022. Its Pro-Am Research Collaboration facilitates research between professionals and amateurs by registering collaborative projects with interested professional and amateur collaborators.

OAO Global Outreach projects such as Women and Girls in Astronomy, 100 Hours of Astronomy, and Telescopes for All encourage communication of science and critical thinking.

- *"Promote dark skies and the pale blue dot message."*
**Efforts to date:**
IAU100 hosted global activities in February 2020 centered on the Pale Blue Dot message of astronomy for global citizenship.

The OAO devotes the month of May to Dark and Quiet Skies activities, with many related events planned.

A collaboration with LEGO in summer 2024 encourages children to look up at the night sky and imagine their own fun constellations ("Funstellations"), for submission and a certificate of participation.

**GOAL 5 ACTIONS**:

Goal 5 primarily pertains to the Office of Astronomy for Education (OAE). In November 2019, the IAU, in partnership with the Max Planck Institute at the Haus der Astronomie in Heidelberg and with the support of the Klaus Tschira Foundation and the Carl Zeiss Foundation, established the Office of Astronomy for Education as the first step of Goal 5. The OAE has its own objectives, which complement the goals of the IAU. An external review of the OAE was completed in Fall 2023, with a recommendation for the OAE to continue.

In the context of the IAU Strategic Plan for 2020-2030, the OAE should:

- *"establish a network of National Astronomy Education Coordinators (NAECs)"*
  **Efforts to date:**
  Since its start of operations, the OAE has quickly built a network of 414 NAECs in 121 countries and territories to date. Community building within this network is an OAE priority, and proceeds through regular meetings (including "1 on 1 dialogues" between a specific country's NAEC team and the OAE") and workshops offered to the NAECs (most recently on editing Wikipedia and on evaluating activities). The NAECs are also actively involved in the Shaw-IAU Workshops and Teacher Training Program (TTP).

- *"analyse the use of astronomy in teaching in IAU countries and identify accessible materials and astronomy literacy guidelines"*
  **Efforts to date:**
  With the help of the NAECs, the OAE has compiled structured descriptions of the role of astronomy in 90 countries and territories. A systematic, survey-based follow-up study is in preparation. Furthermore, the OAE provides a multilingual glossary for ease of adapting resources, and is building a collection of basic illustrations for teaching. It hosts the Big Ideas in Astronomy (a proposed definition of astronomy literacy) as well as the astroEDU portal for peer-reviewed classroom activities, all in the service of identifying a "basic set" of accessible, high-quality resources that can be used for teaching.

- *"encourage standards for teacher training activities"*
  **Efforts to date:**
  Based on a survey of the community, the OAE completed a pilot study on teacher training standards in 2021. The resulting standards will be discussed at the 2024 GA, and subsequently applied to all OAE training workshops. The OAE's Evaluation Toolkit encourages systematic evaluations of educational activities.

- *"organise an annual International School for Astronomy Education (ISAE)"*
  **Efforts to date:**
  The OAE has organised annual "Shaw-IAU Workshops on Astronomy for Education", four of which have been held as large international online schools for astronomy education. Talks are available afterwards on YouTube and in the form of proceedings (e.g. for the 3rd Shaw-IAU Workshop), allowing stakeholders to get up to date on various aspects of astronomy education. The online workshops are complemented by six international in-person Shaw-IAU Workshops, the first of which was held in October 2023 in South Africa, and by local teacher training events that focus on K-12 education, funded by the IAU and Centers/Nodes via the OAE's Teacher Training Program (TTP).

- *"build a database of volunteer IAU members."*
  **Efforts to date:**
  The OAE has created a volunteer mailing list for IAU members and others who have participated in glossary, translation and peer-review work, and sends IAU members calls for participation whenever new volunteer opportunities arise. There are eight [OAE Centers and OAE](#) Nodes collaborating on and supporting the OAE mission.

## III. IAU100 CELEBRATIONS

The IAU celebrated its 100th anniversary in 2019 (officially kicking off at the GA in 2018), spreading its theme of "[Under One Sky](#)" across the globe. Efforts were coordinated through a central office in Leiden and assisted by 123 National Committees. Through intense planning and fundraising (346kEuro raised, plus ~200kEuros for in-kind contributions), IAU100 achieved over 5000 registered activities in 143 countries, featuring the direct involvement of 5-10 million people and a total reach of over 100 million people. All of these remarkable efforts tied in to the goals of the IAU Strategic Plan.

## IV. FUNDRAISING

As noted in the Strategic Plan, the IAU particularly needs more fundraising to support its goals. The IAU is deeply grateful to the foundations below for their recent external support. Some of the grants are up for renewal, and some may be terminated.

- Norwegian Academy of Science and Letters
    - Support for the [IYSA](#) program
    - Support for Young Astronomers lunch at GA
- Gruber Foundation
    - Support for The Gruber Fellowship ([TGF](#)) postdoctoral awards
    - Support for the Gruber prizewinner to speak at the GA
- Kavli Foundation
    - Support for Kavli-IAU [Workshop](#) on Global Coordination, one per triennium
    Support for Kavli-IAU Multidisciplinary [Symposium](#)
    - Support for Young Astronomers Lunch at GA2024
- Shaw Foundation
    - Support for Shaw-IAU Education [Workshops](#) through OAE (through 2025)
    - IAU must develop and propose a new research-related opportunity in Fall 2024 for their consideration
- Gordon and Betty Moore Foundation
    - [I-HOW](#) – workshops for early career astronomers from less-developed regions
- Heising-Simons Foundation

- o [Women and Girls in Astronomy](#) for NA-ROAD - NA-ROAD invited to apply for new grant
  - o Junior Astronomers in US to travel to IAU GA2022 and 2024
- Visegrad Fund
  - o The IAU-International [Visegrad Mobility Fund](#) supports student internships in V4 countries (Hungary, Poland, Czech Republic and Slovakia) for students from Central Europe – ended in 2024
- U.S. IAU [National Committee](#)
  - o Support for the Young Astronomers luncheon and Women in Astronomy luncheon at the GA
  - o Support for the U.S. reception at the GA

In addition, considerable fundraising has been done in various regions. The OAE has raised funds through various German foundations, and the Italian Regional Education Center raised Italian funds. The OAD has generated much regional African support, particularly leading up to GA2024, and the OAO also has additional support in Japan.

The IAU recognizes that continued funding of Office projects and IAU global efforts will require substantial fundraising efforts in the next triennium both from IAU headquarters and from regional appeals by the Offices through their Centers, Nodes, and Regional Offices.

## V. CONCLUSION

This brief review underscores that the IAU, through its members, scientific bodies, and Offices, has worked tirelessly and successfully to achieve many of its stated goals in the IAU Strategic Plan for 2020-2030. It is noteworthy that so much has been accomplished to date, even during the difficulties of the main pandemic years. At 105 years old, the IAU has proven to be a reliable international organization for the stewardship and communication of astronomical discovery, as we all live and work together under one sky. We look forward to the ongoing efforts of our community as the IAU continues in its second century.